\documentclass[11pt,]{article}
\usepackage{authblk}
\usepackage{fullpage}
\usepackage{amssymb,amsmath,mathtools}
\usepackage[T1]{fontenc}
\usepackage{siunitx}
\usepackage[version=3]{mhchem}
\usepackage{xr}
\usepackage{bm,bbm}
\usepackage{pdfpages}
\usepackage{float}
\usepackage{graphicx}
\usepackage{dcolumn}
\usepackage{bm}
\usepackage{color}
\usepackage{tikz}
\usepackage{setspace}
\usepackage{cancel}
\usepackage{appendix}
\usepackage[center,font=small,labelfont=bf]{caption}
\usepackage{float}
\usepackage{braket}
\usepackage{array}
\usepackage{booktabs}
\usepackage{enumitem}
\usepackage{caption}

\usepackage[unicode=true]{hyperref}
\hypersetup{colorlinks=true,
            citecolor=blue,
            linkcolor=blue,
            urlcolor=blue}
            
\usepackage[square,numbers,sort&compress]{natbib}
\usepackage{setspace}

\numberwithin{equation}{section}

\usepackage[left]{lineno}

\newcommand*\linenomathpatch[1]{%
  \cspreto{#1}{\linenomath}%
  \csappto{end#1}{\endlinenomath}%
  \csappto{end#1*}{\endlinenomath}%
}

\linenomathpatch{equation}
\linenomathpatch{gather}
\linenomathpatch{multline}
\linenomathpatch{align}
\linenomathpatch{alignat}
\linenomathpatch{flalign}

\floatname{algorithm}{Algoritmo}

\makeatletter

\newcommand{\hypgeo}[2]{%
  {\vphantom{F}}_{#1}\kern-\scriptspace F_{#2}%
}

\newcommand\approxsim{\mathchoice
  {\@approxsim {\displaystyle}      {1ex} }
  {\@approxsim {\textstyle}         {1ex} }
  {\@approxsim {\scriptstyle}       {.7ex}}
  {\@approxsim {\scriptscriptstyle} {.5ex}}}
\newcommand\@approxsim[2]{%
  \mathrel{%
    \ooalign{%
      $\m@th#1\sim$\cr
      \hidewidth$\m@th#1.$\hidewidth\cr
      \hidewidth\raise #2 \hbox{$\m@th#1.$}\hidewidth\cr
  \input{Reports/reportV0}  }%
  }%
}

\newcommand{\bo}{\raise-1mm\hbox{\Large$\Box$}}

\begin{document}

\title{Self-organized vegetation patterns promote persistence of plant–pollinator mutualisms under environmental stress}

\author[1,2]{Matheus Bongestab}
\author[1]{David Pinto-Ramos}
\author[1,3,4,*]{Ricardo Martinez-Garcia}
\date{}

\affil[1]{\small Center for Advanced Systems Understanding (CASUS); Helmholtz-Zentrum Dresden-Rossendorf (HZDR), Görlitz, Germany}
\affil[2]{Departamento de Física, Universidade Federal da Paraíba, 58051-970 João Pessoa, PB, Brazil}
\affil[3]{ICTP South American Institute for Fundamental Research \& Instituto de F\'isica Te\'orica, Universidade
Estadual Paulista - UNESP, São Paulo SP, Brazil}
\affil[4]{Department of Ecology, Institute of Biosciences, University of S\~ao Paulo, S\~ao Paulo, Brazil}
 \affil[*]{Corresponding author: r.martinez-garcia@hzdr.de}
\maketitle

\begin{abstract}
Mutualisms are key for structuring ecological communities, but they are sensitive to environmental change and fluctuations in population size. Consequently, how mutualisms achieve stability remains an open question in ecological theory. Motivated by previous results in competitive and predator–prey interactions, and by empirical evidence that plant aggregation simultaneously enhances pollinator visitation and intensifies intraspecific competition, we hypothesize that self-organized pattern formation can act as a key stabilizing mechanism of mutualistic interactions. We test this hypothesis using a two-species reaction–diffusion model of a plant–pollinator system that incorporates non-local plant competition and local mutualistic interactions. We first perform a linear stability analysis to determine the conditions under which non-local competition can trigger vegetation pattern formation. We then compute the bifurcation diagrams for both spatial and homogeneous solutions and find that pattern formation enables coexistence at mutualistic strengths below the threshold required in well-mixed populations. This stability gain increases as environmental conditions worsen, because local maxima in vegetation density create the conditions for community persistence despite globally harsh conditions. Moreover, in the strong mutualism limit, the spatial system exhibits multistability between patterned and homogeneous solutions, creating alternative stable configurations that can buffer against fluctuations in population abundance. Spatial self-organization thus stabilizes mutualistic communities through spatial patterns, potentially driving plant-pollinator persistence in stressed environments, including arid ecosystems.
\end{abstract}

\section*{Author Summary}
Mutualistic interactions, such as those between plants and pollinators, are essential for maintaining biodiversity. However, how these interactions remain stable under changing environmental conditions and species abundances is still unclear. Recent empirical work suggests that plant aggregation can have opposing effects on mutualistic stability because vegetation patches attract higher pollinator densities but also intensify intraspecific competition. Yet, a theoretical framework to formalize these effects is still lacking. We introduce a simple spatial model of a plant–pollinator system to investigate how species distributions affect coexistence, and show that plant aggregation has opposing community-level effects: it reduces population sizes but lowers the mutualism threshold for stable coexistence and generates multiple alternative stable states, potentially enhancing resilience to external perturbations. Spatial self-organization therefore provides a local stabilizing mechanism for mutualistic communities that could be particularly relevant in stressed environments.

\section{Introduction}\label{sec:intro}

Mutualisms, in which two species benefit reciprocally from their interaction, are key to the organization of ecological communities \cite{Boucher1982,Stone2020,AraujoLurgi2025,Godoy2026}. Examples include plant-pollinator networks \cite{Bascompte2007,Bascompte2003}, coral-algae symbioses \cite{Stanley2006}, and mycorrhizal networks \cite{Schuessler2011,Simard2012}. Despite their ubiquity and ecological relevance, density-dependent interactions, sensitivity to environmental conditions, and other factors make mutualisms dynamically fragile. This fragility is particularly relevant for obligate mutualisms where at least one species relies on the partner for survival, and a minimum interaction strength is necessary to avoid extinction \cite{hale2021ecological}. In these scenarios, fluctuations in population size and shifts in environmental conditions can change cost-benefit interaction ratios, destabilizing the interaction and potentially leading to the collapse of the community \cite{Sachs2006,toby2010mutualisms,traveset2014mutualistic}. Therefore, understanding the mechanisms responsible for the persistence and stability of mutualistic interactions remains a central question in theoretical ecology. 

Spatial structure is a key driver of population dynamics and can significantly alter the outcomes of species interactions predicted by non-spatial frameworks \cite{Hastings1990,Bolker1999,Turnbull2007,Akhtar2026}. Environmental heterogeneity, dispersal, and local interactions can induce different types of spatial patterns that feed back on population dynamics and determine stability \cite{Cantrell1991, Rietkerk2021, Van_nes2005,samanta2021deterministic,Kastner2025,PintoRamos2025, PintoRamos2026}. For example, dispersal networks between geographically distant patches can promote host-parasitoid coexistence by asynchronous local dynamics \cite{hassell1991spatial}. Likewise, finite-range interactions lead to spatial segregation in competitive and predator-prey systems, which reduces interspecific interaction rates and ultimately enables coexistence \cite{Detto2016,Maciel2021,Simoy2023}. Spatial structure, however, can also have the opposite effect, reducing population size \cite{Plank2025} and even leading to competitive exclusion in scenarios where spatially uniform populations would coexist stably \cite{Bolker2003,Menezes2025}.

In mutualistic systems, spatial processes and features also shape population dynamics and stability in nontrivial ways. Dispersal, foraging, and perceptual ranges determine how often partners encounter and interact with one another, and habitat heterogeneity determines the strength and spatial extent of these interactions \cite{MartinezGarcia2020,amarasekare2004spatial,revilla2016pollinator}. Dispersal between distant patches can introduce tipping points and hysteresis loops at a regional scale, even when isolated patches do not exhibit such thresholds locally  \cite{Denk2024}. In plant-pollinator systems, plant aggregation can attract higher pollinator densities but also intensify intraspecific plant competition \cite{hurtado2023plant}. Although these non-trivial relationships between spatial structure and the organization of mutualistic communities are increasingly acknowledged in large communities, two-species systems have gained significantly less attention. Consequently, the role of spatial pattern formation in stabilizing pairwise mutualistic interactions remains to be studied.

We address this gap using a simple two-species model of a plant-pollinator mutualism. We motivate our model in arid and semi-arid landscapes,
where vegetation density can self-organize into periodic spatial patterns due to increased competition for a limiting resource \cite{martinez2013garcia, MartinezGarcia2023, Jelle2026}. These patterns are common in drylands worldwide \cite{deblauwe2008global,Kastner2024,PintoRamos2026global}, and often dominated by a few species. For example, in the African Sahel and the Australian outback's vegetation patterns, a few species from the \textit{Acacia} genus\textemdash\textit{A. mellifera}, \textit{A. aneura}, \textit{A. bussei}\textemdash, \textit{Combretum micranthum}, and \textit{Guiera senegalensis} account for up to 90\% of the vegetation biomass \cite{deblauwe2008global}. These dominant species rely heavily on insect pollination for reproductive success, with studies reporting strong reductions in seed production in pollinator absence \cite{stone2003pollination, golubov1999demography, boffa1999agroforestry}. Therefore, plant-pollinator mutualism is a key ecological interaction in these systems.

We focus on an obligate-facultative system where the mutualism is obligate for plants but pollinators can persist independently \cite{ollerton2011many,bronstein1994our}. Our results show that, in these scenarios, vegetation patterns trigger the aggregation of its pollinator partner and, although they reduce population densities relative to uniform distribution when mutualism is strong, they allow both species to persist under weaker mutualistic interactions than predicted by the non-spatial model. Moreover, this stability gain increases with environmental harshness because spatial patterning keeps population densities locally above the threshold that defines the obligate mutualism. Spatial self-organization thus provides a local stabilizing mechanism for mutualistic communities that could be particularly relevant for understanding the persistence of plant-pollinator interactions in stressed environments.

\section{Materials and Methods}
\label{sec:met}

We propose a spatial model describing a mutualistic plant-pollinator interaction. Because we are interested in investigating the role of spatial patterning on the stability and persistence of the mutualistic interaction, we consider a simple vegetation model known to produce regular spatial patterns \cite{martinez2013garcia}, \begin{equation}\label{eq:veg}
    \frac{\partial V(X,T)}{\partial T} = s\,P_\mathrm{E}(c,\tilde{V}) V\,P\left(1-\frac{V}{K}\right) - d\,V+D_V\frac{\partial^2 V}{\partial X^2},
\end{equation}
where $V(X,T)$ is the vegetation biomass density field, the diffusion term accounts for short-range dispersal, and $d$ is the baseline mortality rate. The first term on the right side accounts for vegetation growth, which occurs at a maximum rate $s$ through interaction with pollinators at density $P(X,T)$. This growth is inhibited by two competitive interactions: local competition for space via the growth-limiting linear term with carrying capacity $K$, and long-range competition for a limiting resource via an establishment probability $P_\mathrm{E}(c,\tilde{V})\in [0,1]$. This establishment probability is a monotonically decreasing function of the biomass density averaged within a neighborhood centered at $X$, $\tilde{V}(X,T)$, and a positive parameter $c$ weighting the interaction strength. Following previous work \cite{martinez2013garcia,martinez2014garcia}, and to facilitate our analyses, we use 
\begin{equation}\label{eq:pe}
    P_\mathrm{E}(c,\tilde{V}) = \frac{1}{1+c\,\tilde{V}},
\end{equation}
where the non-local vegetation biomass density $\tilde{V}$
 \begin{equation}\label{eq:nonlocal_dens}
   \tilde{V}(X,T) = \frac{1}{2R}\int_{X-R}^{X+R} V(X',T)dX'
\end{equation}
    already assumes a top-hat kernel of lateral length $R$, which accounts for finite-range competition among plants mediated by their root systems \cite{Lefever1997,Borgogno2009,MartinezGarcia2023}. The top-hat kernel is the simplest choice leading to spatial patterns, which makes it a standard choice in models of non-local competition due to its mathematical and computational tractability \cite{Hernandez-Garcia2004,martinez2013garcia,Piva2021}. This choice is not restrictive, and our results remain qualitatively unchanged as long as a kernel function that yields spatial patterns is chosen. This occurs for a broad class of kernels, including platykurtic generalized exponentials and functions with finite support \cite{Lopez2004,martinez2014garcia,Surendran2025,Jelle2026}. A more realistic description of root-mediated competition would account for the complex responses of root systems to resource availability and root density \cite{Cabal2020,Cabal2024}, but incorporating such processes into the nonlocal term is beyond the scope of this work and remains an interesting direction for future study.

The pollinator's spatial dynamics is given by 
\begin{equation}\label{eq:pol}
    \frac{\partial P(X,T)}{\partial T} = rP - l P^2 + m P V + D_P\frac{\partial^2 P}{\partial X^2},
\end{equation}
such that pollinators have a net growth rate $r$ limited by intraspecific competition with intensity $l$ and enhanced by a mutualistic interaction with plants with strength $m$. The diffusion term in this case provides a simple description for pollinator movement and hence $D_P\gg D_V$. All model parameters are positive except the pollinator net growth rate, which is negative when baseline mortality is higher than reproduction (see Table~\ref{tb:qnt} for a parameter summary).
\begin{table}[t]
    \centering
    \begin{tabular}{||c|c||} 
        \hline
        Symbol & Parameter meaning \\ [0.5ex] 
        \hline\hline
        $s$& Strength of mutualistic interaction on vegetation growth \\ 
        \hline
        $d$& Vegetation death rate  \\
        \hline
        $c$& Strength of non-local competition \\
        \hline
        $K$& Carrying capacity emerging due to competition for space \\
        \hline
        $r$& Pollinator net growth rate \\ 
        \hline
        $l$& Intensity of pollinator intraspecific competition\\
        \hline
        $m$&  Strength of mutualistic interaction on pollinator growth \\ 
        \hline
        $R$& Range of non-local competition \\ [0.5ex]
        \hline
    \end{tabular}
    \caption{Summary of model parameters and their ecological interpretation.}
    \label{tb:qnt}
\end{table}

To facilitate the model analysis, we scale space, time, and population densities as
\begin{align*}
    V = Kv; \quad
    P = \frac{d}{s}\,p; \quad
    T = \frac{1}{d} \, t; \quad
    X = Rx,
\end{align*}
which allows us to write the model equations in terms of dimensionless quantities as
\begin{eqnarray}   
        \dfrac{\partial v(x, t)}{\partial t} &=& \dfrac{(1 - v)v p}{1 + \alpha \tilde{v}} - v + \overline{D}_v\dfrac{\partial^2 v}{\partial x^2}, \label{eq:spatialveg}\\
    \dfrac{\partial p(x, t)}{\partial t} &=& \beta p - \gamma p^2 + \mu v p + \overline{D}_p\dfrac{\partial^2 p}{\partial x^2},\label{eq:spatialpol}
\end{eqnarray}
where the averaged density is
 \begin{equation}\label{eq:nonlocal_dens-app}
   \tilde{v}(x,t) = \frac{1}{2}\int_{x-1}^{x+1} v(x',t)dx'.
\end{equation}
The new parameters are defined in terms of the old ones as $\alpha \equiv Kc$,  $\beta \equiv r/d$, $\gamma \equiv l/s$, $\mu \equiv Km/d$, $\overline{D}_v \equiv D_V/(dR^2)$, $\overline{D}_p \equiv D_P/(dR^2)$. 

In the next sections, we analyze this model using a combination of mathematical and numerical approaches. All numerical simulations were performed using an Euler method in a one-dimensional array with $N_x=200$ grid points and step size $dx=0.05$, making a system size $l=10R$. We used periodic boundary conditions and ran each simulation for a maximum time $t_f=10^3$ with time discretization $dt=0.01$ unless otherwise specified. At this time, all simulations converged to an equilibrium state.

\section{Results}
\label{sec:res}
\subsection{Stability of the spatially uniform steady states}\label{sec:homo}

We first analyze the plant-pollinator system in the non-spatial limit to obtain the homogeneous solutions for each population. In this limit, the spatial average $\Tilde{v}$ becomes the mean-field average $v$, so the establishment probability is $P_E(c,v)$. We follow the standard procedure, obtaining the fixed points as the intersections of each population's nullclines, and determining their stability from the eigenvalues of the model's Jacobian matrix evaluated at each fixed point. 
To obtain the system nullclines we impose $dv/dt = 0$ and $dp/dt=0$ in Eqs.\, \eqref{eq:spatialveg}-\eqref{eq:spatialpol}, respectively. The system has trivial nullclines for the pollinators ($p=0$) and the vegetation ($v=0$), as well as a non-trivial nullcline for the pollinators 
\begin{equation}
p=\bar\beta + \bar\mu v, \label{null_1}
\end{equation}
and one for the vegetation
\begin{equation}
p=\frac{1+\alpha v}{1-v}. \label{null_2}
\end{equation}
Notice that we have also defined $\bar\beta=\beta/\gamma$ and $\bar\mu=\mu/\gamma$. Linearizing Eqs.\,\eqref{eq:spatialveg}-\eqref{eq:spatialpol} around an arbitrary fixed point $(v^*,\,p^*)$, we obtain the following Jacobian matrix,
\begin{equation}
\mathbb{J}=
\begin{bmatrix}
    \dfrac{(1-2v)p}{1+\alpha v} - \dfrac{\alpha (1-v)vp}{(1+\alpha v)^2} - 1 & \dfrac{1}{1+\alpha v}(1 - v)v \\
    \bar\mu p & \bar\beta - 2p + \bar\mu v
\end{bmatrix}_{(v^*,\,p^*)}.
\label{eq:jac}
\end{equation}

We find that our plant-pollinator model may have up to four homogeneous stationary solutions. The first one is a trivial state in which both populations go extinct. For this solution, the eigenvalues of the Jacobian matrix are $\lambda_1=-1$ and $\lambda_2=\bar\beta$, so the fixed point is stable when $\bar\beta<0$ and unstable (saddle) otherwise. The second one, $(v^*, p^*) = (0, \bar\beta)$ represents a non-vegetated state with pollinators, and it exists provided that the pollinator growth rate is positive $\bar\beta>0$ as a result of the trivial solution becoming unstable. The Jacobian eigenvalues are $\lambda_1=-\bar\beta$ and $\lambda_2=\bar\beta-1$, indicating that this solution is stable when $0<\bar\beta\leq 1$ and unstable (saddle) for $\bar\beta>1$. Finally, there are up to two additional fixed points where both species coexist. We can obtain these fixed points analytically as the intersection points between the nullclines in Eqs.\,\eqref{null_1} and \eqref{null_2} where both plant and pollinator densities are positive,
\begin{eqnarray}
    v=\frac{-(\alpha + \bar{\beta}-\bar{\mu})\pm\sqrt{(\alpha + \bar{\beta}-\bar{\mu})^2-4\bar{\mu}(1-\bar{\beta})}}{2\bar{\mu}},\nonumber \\
    p=\frac{\bar{\beta}}{2}-\frac{(\alpha-\bar{\mu})}{2} \pm\frac{\sqrt{(\alpha + \bar{\beta}-\bar{\mu})^2-4\bar{\mu}(1-\bar{\beta})}}{2}.\label{eq:coexistence}
\end{eqnarray}

Alternatively, we can analyze the behavior of the nullclines to understand the existence and stability of these fixed points. We first consider the case where $\bar\beta<1$. In this regime, the linear pollinator nullcline takes smaller values than the vegetation nullcline when $v=0$. $\alpha$ and $\bar\mu$, which control the derivative of these nullclines, will determine whether they intersect twice or zero times. Specifically, for $\alpha$ small enough or $\bar\mu$ large enough, both fixed points exist, one being stable and the other an unstable saddle point. In this regime, the plant-pollinator community exhibits alternative stable states. One of these alternative stable states is always the coexistence of both species, while the second shifts from community collapse when $\bar\beta < 0$ to vegetation extinction and pollinator persistence when $\bar\beta > 0$. Finally, when $\bar\beta>1$, there is always one coexistence fixed point that is stable for any value of $\alpha$ and $\mu$.

\begin{figure}[t]
    \centering
    \includegraphics[width=0.9\textwidth]{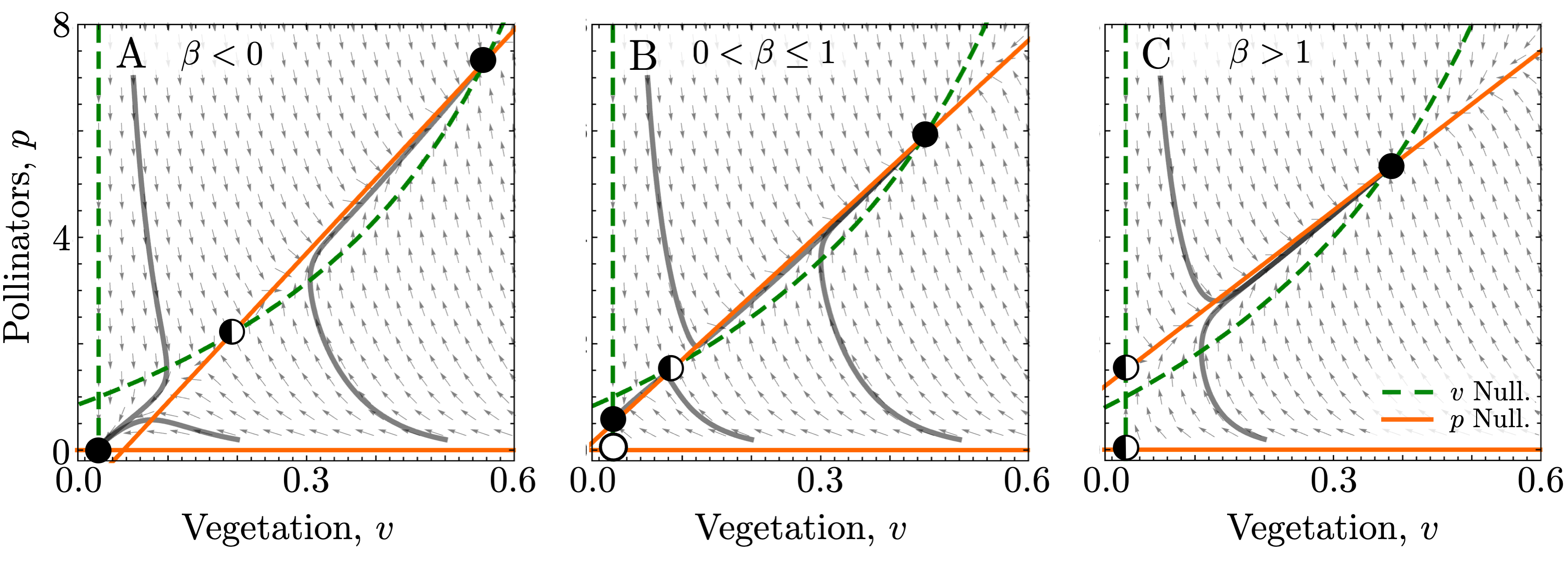}
    \caption{Phase space of the non-spatial model for the possible regimes: A) obligate mutualism for plants and pollinators and bistability between community collapse and coexistence, $\bar\beta=-0.5$; B) obligate mutualism only for plants and bistability between coexistence and plant extinction, $\bar\beta=0.5$; C) obligate mutualism only for plants and monostable coexistence, $\bar\beta=2.0$. Other parameters: $\alpha=5.0$, $\bar\mu = 9.0,\; 13.0,\; 15.0$. Grey curves are examples of solution trajectories. Filled circles indicate stable fixed points and semi-circles indicate unstable saddle nodes.}
    \label{fig:1}
\end{figure}

Therefore, the ratio between pollinator net growth rate and vegetation death, $\bar\beta$, defines three scenarios in the phase space (Fig.\,\ref{fig:1}). In the first case, the pollinator's net growth rate is negative, $\bar\beta<0$, and both species can survive only if mutualism is strong enough relative to intraspecific competition. When this condition is met, a stable coexistence fixed point exists together with the extinction state, and survival requires initial densities within the basin of attraction of the former (Fig.\,\ref{fig:1}A). However, below a minimum mutualism strength, the coexistence fixed point disappears via a saddle-node bifurcation, and extinction becomes the only stable state. Mutualism is thus obligate for both species when $\bar\beta<0$, and a sufficiently strong mutualism is a necessary, but not sufficient, condition for survival. In the second scenario, the pollinators' net growth rate is positive but smaller than the vegetation's death, $0 < \bar\beta \leq 1$. This scenario allows pollinators to persist in the absence of plants, for which the mutualism is still obligate (Fig.\,\ref{fig:1}B). In these two scenarios, the community exhibits alternative states, and species coexistence depends on initial population sizes. Finally, when pollinators grow faster than vegetation dies, $\bar\beta > 1$ (Fig.\,\ref{fig:1}C), the two species will always coexist regardless of the intensity of intraspecific competition and the strength of the mutualism. In this scenario, the high growth rate of the pollinators outbalances any population loss, enabling coexistence, while the strength of mutualism and intraspecific competition only determines population abundances at equilibrium. Therefore, the community does not exhibit alternative stable states, and varying parameters (other than $\bar{\beta}$) cannot trigger community tipping.

\subsection{Pattern formation instability}
Next, we analyze the fully spatial model to obtain the conditions for Turing instabilities \cite{turing1952chemical,murray2007mathematical}. 
Because we are interested in how vegetation spatial patterns influence community tipping points, we focus on parameter regimes where the non-spatial model is bistable, i.e., $\bar{\beta}\leq 1$, which allows for a direct comparison between the spatial and non-spatial models. Additionally, we fix both diffusion coefficients, imposing $\overline{D}_v=10^{-4}\ll\overline{D}_p=10^{-3}$ to account for pollinator movement being faster than plant dispersal. With these assumptions, the model has two free parameters left: the intensity of the mutualistic interaction $\bar\mu$ and plant intraspecific competition $\alpha$.

\begin{figure}[t]
    \centering
    \includegraphics[width=0.9\textwidth]{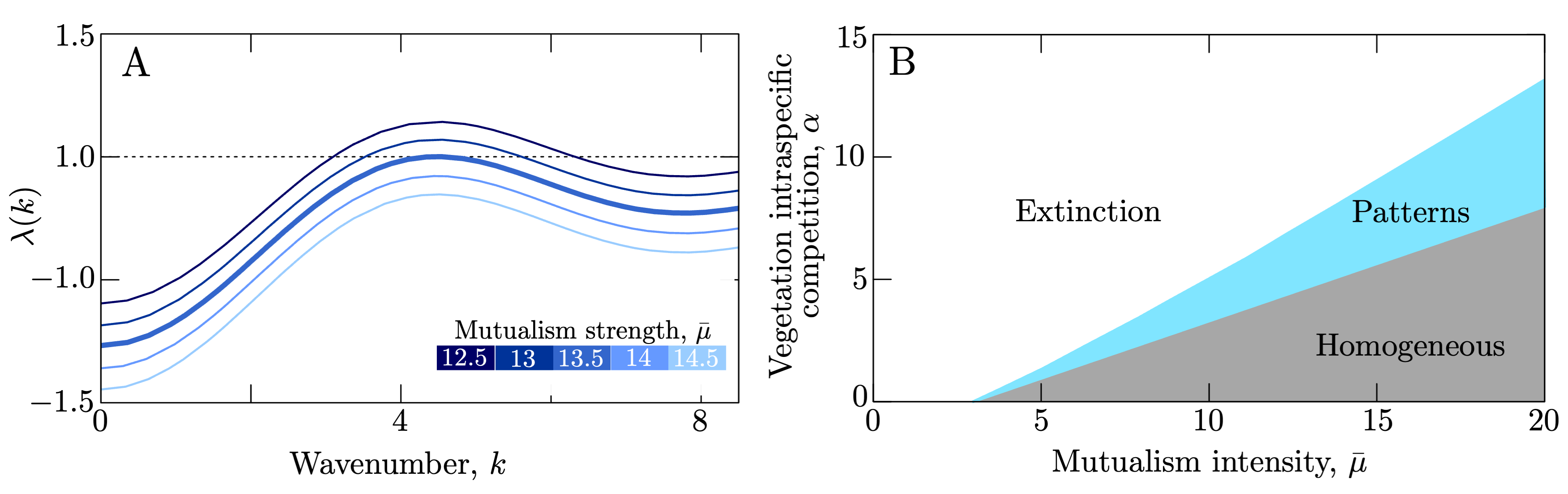}
    \caption{Turing instability in the presence of homogeneous alternative stable states, $\bar{\beta}<1$. A) Real part of the largest eigenvalues at different mutualism strengths. The thickest curve corresponds to the mutualism strength at the Turing instability onset, $\bar\mu^* = 13.5$ for $\alpha=5.0$. B) Onset of the Turing instability (cyan-gray transition) and tipping of self-organized populations (cyan-white transition) in the $\alpha$-$\bar\mu$ parameter space. Other parameters: , $\bar\beta=0.5$}
    \label{fig:2}
\end{figure}

To investigate the possible formation of vegetation patterns, we perform a linear stability analysis around the stable equilibrium that leads to species coexistence. The Jacobian of the spatial system is (see Appendix\,\ref{ap:A} for the full derivation)
\begin{equation}
    \mathcal{A} = \begin{pmatrix}
        p(1-2v)\dfrac{1}{1+\alpha v} - \dfrac{\alpha (1-v)vp}{(1+\alpha v)^2}\hat{G}(k) - 1 - \overline{D}_v k^2 \; & \; v(1-v)\dfrac{1}{1+\alpha v} \\
        \bar\mu p & \bar\beta - 2 p + \bar\mu v - \overline{D}_p k^2
    \end{pmatrix}_{(v,p)^*}
\end{equation}
where $\hat{G}(k)=\sin(k)/k$ is the Fourier transform of the top-hat kernel $G(x-x')$ and the subindex $(v,p)^*$ indicates that the population densities in the matrix are evaluated at the coexistence fixed point. The eigenvalues of this matrix are
\begin{equation*}
        \lambda_{1,2}(k) = \dfrac{1}{2}\left(\mathrm{tr}(\mathcal{A}) \pm \sqrt{\mathrm{tr}^2(\mathcal{A}) - 4\,\mathrm{det}(\mathcal{A})}\right).
        \label{eq:lambda}
\end{equation*}
\noindent In general, the homogeneous state $(v,p)^*$ is unstable to heterogeneous perturbations if the maximum of the real part of the largest eigenvalue is positive at a non-zero wavenumber $k_c\neq0$. This instability causes periodic solutions to grow exponentially, leading to a spatial pattern once the nonlinearities saturate the growth. The two free parameters favor the formation of spatial patterns when their changes lead to worsening conditions for vegetation and pollinator growth. Consequently, when $\bar\mu$ is high and $\alpha$ is low, representing high mutualism and low intraspecific vegetation competition, the spatially uniform population distributions are stable and patterns do not form. However, as $\bar\mu$ decreases or $\alpha$ increases, the system crosses a Turing instability, allowing self-organized patterns to form (Fig.\,\ref{fig:2}A,\,B). Eventually, if these two parameters continue to change in the direction of worsening growth conditions, the system suffers an abrupt transition from the pattern state to the remaining homogeneous state: community collapse if $\bar{\beta}<0$ or pollinator-only survival if $0<\bar\beta<1$ (Fig.\,\ref{fig:2}B). Therefore, self-organized patterns emerge in response to worsening growth conditions, similarly to what happens in models that consider only vegetation spatial dynamics in arid and semi-arid systems \cite{martinez2013garcia}. 

\subsection{Effect of spatial patterns on community stability}
\begin{figure}[t]
    \centering
    \includegraphics[width=0.8\linewidth]{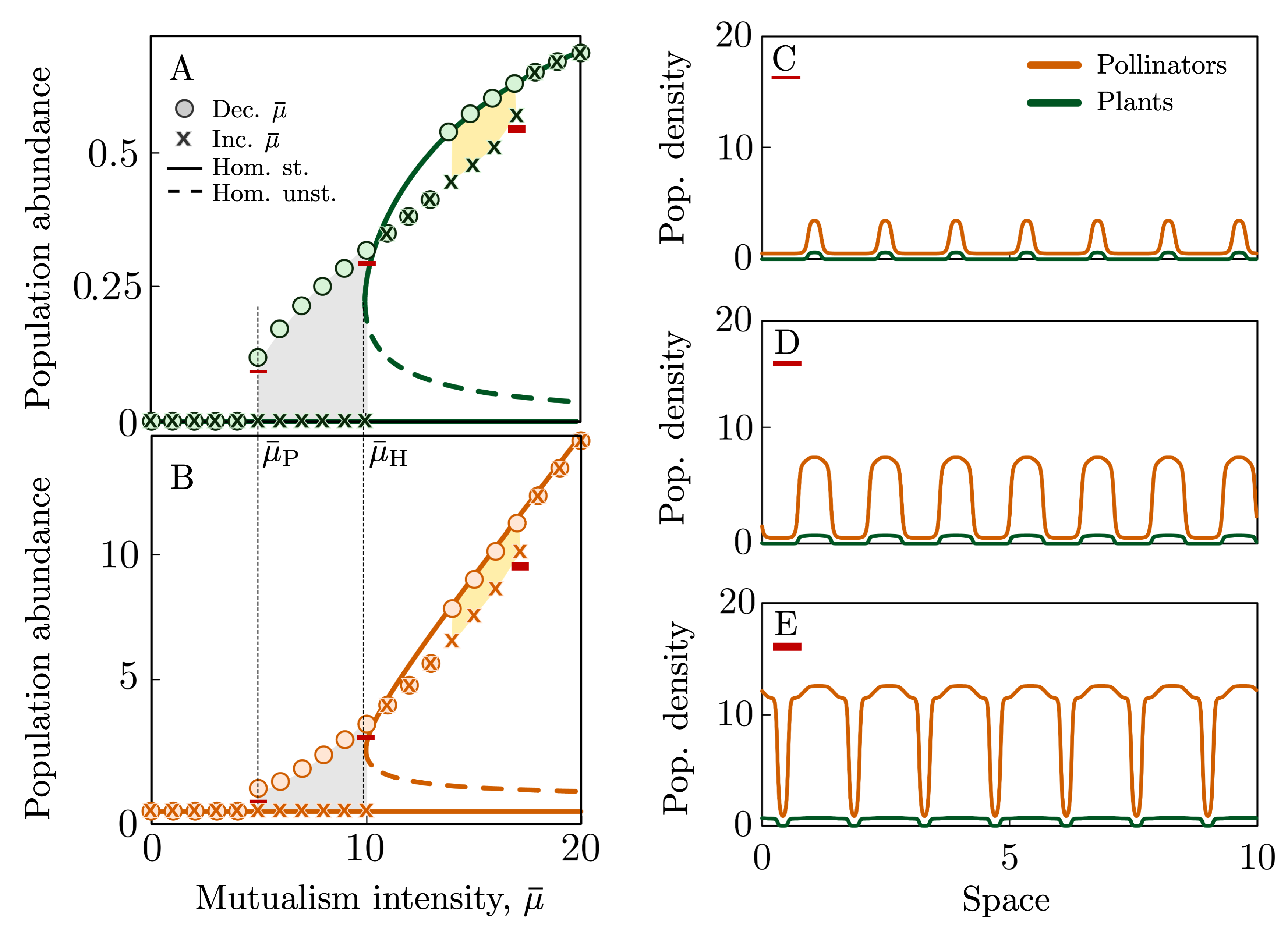}
    \caption{A, B) Plant (green) and pollinator (orange) bifurcation diagrams. Curves correspond to the non-spatial model, with solid and dashed lines representing stable and unstable steady states, respectively. Symbols correspond to numerical simulations of the spatial model. Circles are obtained starting at $\bar\mu=20$ and reducing the mutualism strength quasi-adiabatically, whereas crosses are obtained by increasing $\bar\mu$ quasi-adiabatically starting at $\bar\mu=11$. The gray and yellow shadow highlight the two hysteresis loops, and the thin vertical dashed lines limit the gain in mutualism stability due to spatial patterns, $\bar\mu_\mathrm{H}$ and $\bar\mu_\mathrm{P}$. The dark-red lines indicate the points for which spatial patterns are shown in C-E.  C-E) Spatial patterns of population density at different values of the mutualism strength: C) $\bar\mu=6$; D) $\bar\mu=11$; E) $\bar\mu=17$ (lower pattern branch). Other parameter values: $\alpha = 5.0$ and $\bar\beta=0.5$}
    \label{fig:4}
\end{figure}

We next study how pattern formation changes the possible equilibrium states along a gradient of mutualism intensity while keeping the intensity of intraspecific vegetation competition fixed. We focus on the case $0<\bar\beta\leq1$, but similar results hold for $\bar\beta <0$ with different homogeneous equilibrium states (Fig. \ref{fig:1}A, B). At high $\bar\mu$, the system does not develop patterns and therefore the solution of the spatial model converges to the fixed points of the non-spatial equations (disks and cross fall on the solid curve in Fig.\,\ref{fig:4}A,\,B). As the mutualism intensity decreases, the system crosses the Turing bifurcation point and develops spatial patterns. These patterns form with both populations in phase, and the characteristic cluster size decreases with decreasing $\bar{\mu}$ (Fig.\,\ref{fig:4}C-E). Even though pattern configurations result in lower population sizes than homogeneous ones, these smaller populations can persist beyond the tipping point of the non-spatial system. Therefore, spatial patterns promote community stability by reducing the threshold in the mutualism intensity required for stable coexistence.

Pattern formation also leads to a more complex landscape of alternative stable states across a mutualism-intensity gradient. Close to the onset of pattern formation, spatial patterns can coexist with the two stable homogeneous solutions and the system is multistable. At lower mutualism intensities, because pattern configurations can persist beyond the mutualism threshold that ensures coexistence in homogeneous populations, the system exhibits bistability between a patterned configuration in which both species coexist and the pollinator-only homogeneous state. This diversity of alternative stable states introduces two hysteresis loops in the system: one between the homogeneous and patterned coexistence states (yellow shaded region in Figs.\,\ref{fig:4}A\,,B), and another between the pollinator-only homogeneous patterned coexistence states (gray shaded region in Figs.\,\ref{fig:4}A\,,B)

\begin{figure}[t]
    \centering
    \includegraphics[width=0.5\linewidth]{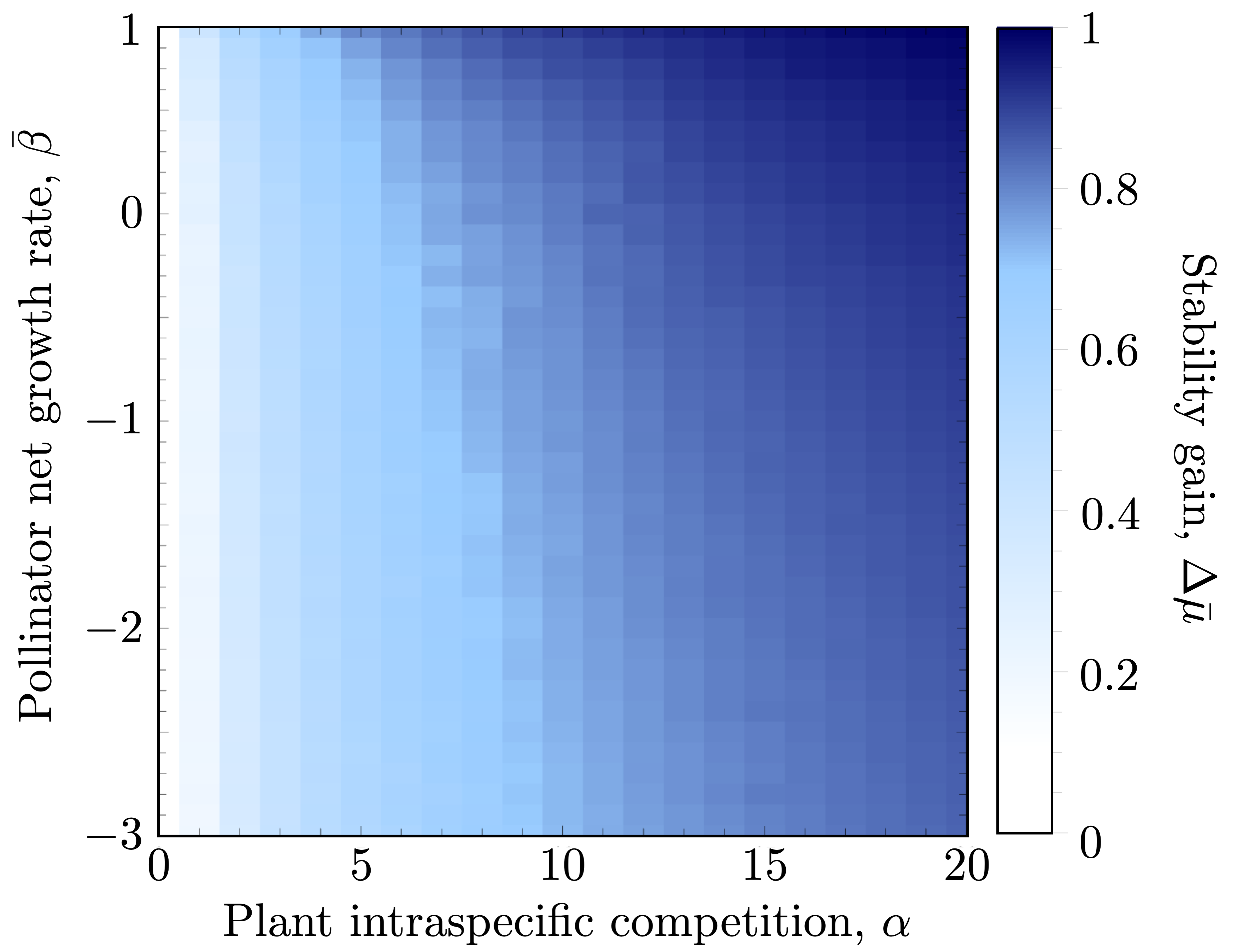}
    \caption{Stability gain, $\Delta \bar\mu$ across worsening environmental conditions for pollinators (decreasing net growth rate $\bar\beta$) and plants (increasing intraspecific competition $\alpha$).}
    \label{fig:5}
\end{figure}

Finally, to quantify how much pattern formation promotes species coexistence, we compare the mutualism intensity at which coexistence becomes unstable and the system shifts to either a pollinator-only or a non-populated state, depending on $\bar{\beta}$. We compare these mutualism-intensity thresholds for the non-spatial model solutions, which assume uniformly distributed populations, and the full spatial solutions, which produce patterned populations. The corresponding critical thresholds, $\bar{\mu}_{\mathrm{H}}$ and $\bar{\mu}_{\mathrm{P}}$ respectively, are the system’s tipping points. We can obtain the non-spatial system's tipping point analytically by imposing that Eq.\,\eqref{eq:coexistence} gives exactly one value,
\begin{eqnarray}
    \bar{\mu}_\text{\tiny H}=2+\alpha-\bar{\beta}+ 2\sqrt{(1-\bar{\beta})(1+\alpha)}.
\end{eqnarray}
$\bar{\mu}_{\mathrm{P}}$ has to be computed numerically because there is no analytical solution for the patterns. Using these values, we define the relative stability gain $\Delta \bar\mu = (\bar{\mu}_{\mathrm{H}} - \bar{\mu}_{\mathrm{P}})/\bar{\mu}_{\mathrm{H}}$. From this definition, it follows that $\Delta \bar\mu=1$ when the stability gain due to spatial patterns is such that both species coexist in the absence of mutualism, whereas $\Delta \bar\mu=0$ when $\bar{\mu}_{\mathrm{H}} = \bar{\mu}_{\mathrm{P}}$. We obtained this quantity across different values of plant intraspecific competition, $\alpha$, and pollinator net growth rate, $\bar\beta$. $\Delta\bar\mu$ tends to a maximum gain for strong non-local intraspecific competition between plants, while $\bar\beta$ has a very weak effect (Fig.\,\ref{fig:5}).

\section{Discussion and Conclusions}

We studied how self-organized spatial patterns influence the stability of mutualistic communities, using a reaction-diffusion model for a two-species plant-pollinator system with obligate mutualism for the vegetation species. In the non-spatial limit, the model exhibits three different regimes at low population densities. Species coexistence is the only possible equilibrium if the pollinator's reproduction rate is greater than the vegetation death times the strength of pollinator intraspecific competition, $\beta/(\gamma d)> 1$. Otherwise, the community exhibits alternative stable states between coexistence and full community collapse when $\beta<0$, or a pollinator-only state. Spatial patterns around the coexistence state affect community properties mainly in three ways: they reduce population sizes at equilibrium, lower the mutualism threshold for stable coexistence, and create a more complex landscape of alternative stable states.

Previous models focused on large systems with many species \cite{Bascompte2007,Bascompte2003}, and investigated the effect of spatial processes at the metacommunity level considering dispersal between multiple patches \cite{amarasekare2004spatial,revilla2016pollinator}. We focused on spatial processes within a single patch, using a simpler model with only two interacting species. While spatial patterning reduces population abundances relative to homogeneous populations when the mutualism is sufficiently intense, it enhances community stability across a wide range of interaction strengths and environmental stress. Specifically, we found a positive feedback between spatial patterning and pollination efficiency that reinforces community stability: vegetation redistribution into patches enhances pollinator effects, which, in turn, favors vegetation growth.

Spatial patterns also result in a more complex landscape of community alternative stable states, including multistability between patterned and non-patterned states. Alternative stable states are common in reaction-diffusion models of vegetation dynamics, and abrupt transitions among them following environmental threshold crossings determine how water-limited ecosystems will respond to increased perturbation and restoration strategies \cite{Van_nes2005,Rietkerk2021,Michaels2024}. 
Our results highlight the importance of accounting for interactions between vegetation and other species in those ecosystems to better understand how they respond to changing environmental conditions. Mutualisms, for example, change the possible stable states and, consequently, their basins of attraction and the environmental thresholds at which changes among them occur.

The richness of alternative stable states in our spatial model indicates that nonlinearities in the species interactions, represented by different functional responses, are key to community stability \cite{AraujoLurgi2025,hale2021ecological}. We considered the simplest functional response, in which the strength of the interaction increases linearly with the density of the partner species. Future work should generalize this modeling choice to account for nonlinear functional responses. In plant-pollinator interactions, for example, pollinator handling times are better described by a saturating Holling type II function, and reduced encounter rates at low vegetation densities by type-III response functions \cite{holling1959components}. Another direction of future research should generalize the pairwise analysis to larger communities, accounting for interaction networks typical of mutualistic communities \cite{hale2021ecological}. Mutualists are likely to interact with many partners, and mutualistic effects are more complex in mixed communities \cite{holland2010consumer,valdovinos2019mutualistic}. Therefore, additional work in this direction would provide a better understanding of how mutualistic network structure effects interact with spatial self-organization in complex ecosystems.

\section*{Code Availability}
The code developed for the analysis presented in this work is available in the \href{https://github.com/athemusb/Mutualism}{GitHub Repository}.

\section*{Acknowledgments}
This work is supported by Coordenação de Aperfeiçoamento de Pessoal de Nível Superior (CAPES), grant no.~88887.688488/2022-00 (MB) and co-financed by the Deutscher Akademischer Austauschdienst (DAAD, grant no.~57710871), under a collaborative short-term research program. This work was partially funded by the Center for Advanced Systems Understanding (CASUS), which is financed by Germany’s Federal Ministry of Education and Research (BMBF) and by the Saxon Ministry for Science, Culture and Tourism (SMWK) with tax funds on the basis of the budget approved by the Saxon State Parliament. RMG was partially supported by the S\~ao Paulo Research Foundation (FAPESP, Brazil) through ICTP-SAIFR grant no.~2021/14335-0 (R.M.-G).

\bibliography{ref}

@article{turing1952chemical,
  title     = {The Chemical Basis of Morphogenesis},
  author    = {Turing, Alan M.},
  journal   = {Philosophical Transactions of the Royal Society of London. Series B, Biological Sciences},
  volume    = {237},
  number    = {641},
  pages     = {37--72},
  year      = {1952},
  publisher = {The Royal Society},
  doi       = {10.1098/rstb.1952.0012}
}

@article{martinez2013garcia,
  title     = {Vegetation Pattern Formation in Semiarid Systems without Facilitative Mechanisms},
  author    = {Mart{\'i}nez-Garc{\'i}a, Ricardo and Calabrese, Justin M. and Hern{\'a}ndez-Garc{\'i}a, Emilio and L{\'o}pez, Crist{\'o}bal},
  journal   = {Geophysical Research Letters},
  volume    = {40},
  number    = {23},
  pages     = {6143--6147},
  year      = {2013},
  publisher = {Wiley Online Library},
  doi       = {10.1002/2013GL058797}
}

@article{martinez2014garcia,
  title     = {Minimal Mechanisms for Vegetation Patterns in Semiarid Regions},
  author    = {Mart{\'i}nez-Garc{\'i}a, Ricardo and Calabrese, Justin M. and Hern{\'a}ndez-Garc{\'i}a, Emilio and L{\'o}pez, Crist{\'o}bal},
  journal   = {Philosophical Transactions of the Royal Society A: Mathematical, Physical and Engineering Sciences},
  volume    = {372},
  number    = {2027},
  pages     = {20140068},
  year      = {2014},
  publisher = {The Royal Society},
  doi       = {10.1098/rsta.2014.0068}
}

@article{hassell1991spatial,
  title     = {Spatial Structure and Chaos in Insect Population Dynamics},
  author    = {Hassell, Michael P. and Comins, Hugh N. and May, Robert M.},
  journal   = {Nature},
  volume    = {353},
  number    = {6341},
  pages     = {255--258},
  year      = {1991},
  publisher = {Nature Publishing Group},
  doi       = {10.1038/353255a0}
}

@article{ollerton2011many,
  title     = {How Many Flowering Plants Are Pollinated by Animals?},
  author    = {Ollerton, Jeff and Winfree, Rachael and Tarrant, Sam},
  journal   = {Oikos},
  volume    = {120},
  number    = {3},
  pages     = {321--326},
  year      = {2011},
  publisher = {Wiley Online Library},
  doi       = {10.1111/j.1600-0706.2010.18644.x}
}

@article{hurtado2023plant,
  title     = {Plant Spatial Aggregation Modulates the Interplay between Plant Competition and Pollinator Attraction with Contrasting Outcomes of Plant Fitness},
  author    = {Hurtado, Mar{\'i}a and Godoy, Oscar and Bartomeus, Ignasi},
  journal   = {Web Ecology},
  volume    = {23},
  number    = {1},
  pages     = {51--69},
  year      = {2023},
  publisher = {Copernicus Publications},
  doi       = {10.5194/we-23-51-2023}
}

@article{valdovinos2019mutualistic,
  title     = {Mutualistic Networks: Moving Closer to a Predictive Theory},
  author    = {Valdovinos, Fernanda S.},
  journal   = {Ecology Letters},
  volume    = {22},
  number    = {9},
  pages     = {1517--1534},
  year      = {2019},
  publisher = {Wiley Online Library},
  doi       = {10.1111/ele.13279}
}

@article{Boucher1982,
  author = {Boucher, D.~H. and James, S. and Keeler, K.~H.},
  title = {The ecology of mutualism},
  journal = {Annual Review of Ecology and Systematics},
  year = {1982},
  volume = {13},
  pages = {315--347},
  doi = {10.1146/annurev.es.13.110182.001531}
}

@article{Bolker2003,
  title     = {Spatial Dynamics in Model Plant Communities: What Do We Really Know?},
  author    = {Bolker, Benjamin M. and Pacala, Stephen W. and Neuhauser, Claudia},
  journal   = {The American Naturalist},
  volume    = {162},
  number    = {2},
  pages     = {135--148},
  year      = {2003},
  publisher = {University of Chicago Press},
  doi       = {10.1086/376575}
}

@article{Bolker1999,
  title     = {Spatial Moment Equations for Plant Competition: Understanding Spatial Strategies and the Advantages of Short Dispersal},
  author    = {Bolker, Benjamin M. and Pacala, Stephen W.},
  journal   = {The American Naturalist},
  volume    = {153},
  number    = {6},
  pages     = {575--602},
  year      = {1999},
  publisher = {University of Chicago Press},
  doi       = {10.1086/303199}
}

@article{Borgogno2009,
  title     = {Mathematical Models of Vegetation Pattern Formation in Ecohydrology},
  author    = {Borgogno, F. and D'Odorico, P. and Laio, F. and Ridolfi, L.},
  journal   = {Reviews of Geophysics},
  volume    = {47},
  number    = {1},
  pages     = {RG1005},
  year      = {2009},
  publisher = {American Geophysical Union},
  doi       = {10.1029/2007RG000256}
}

@article{bronstein1994our,
  title     = {Our Current Understanding of Mutualism},
  author    = {Bronstein, Judith L.},
  journal   = {The Quarterly Review of Biology},
  volume    = {69},
  number    = {1},
  pages     = {31--51},
  year      = {1994},
  publisher = {University of Chicago Press},
  doi       = {10.1086/418432}
}

@article{hale2021ecological,
  title     = {Ecological Theory of Mutualism: Robust Patterns of Stability and Thresholds in Two-Species Population Models},
  author    = {Hale, Kayla R. S. and Valdovinos, Fernanda S.},
  journal   = {Ecology and Evolution},
  volume    = {11},
  number    = {24},
  pages     = {17651--17671},
  year      = {2021},
  publisher = {Wiley Online Library},
  doi       = {10.1002/ece3.8453}
}

@article{amarasekare2004spatial,
  title     = {Spatial Dynamics of Mutualistic Interactions},
  author    = {Amarasekare, Priyanga},
  journal   = {Journal of Animal Ecology},
  volume    = {73},
  number    = {1},
  pages     = {128--142},
  year      = {2004},
  publisher = {Wiley Online Library},
  doi       = {10.1046/j.0021-8790.2004.00788.x}
}

@article{revilla2016pollinator,
  title     = {Pollinator Foraging Adaptation and Coexistence of Competing Plants},
  author    = {Revilla, Tom{\'a}s A. and Kr{\v{z}}ivan, Vlastimil},
  journal   = {PLOS ONE},
  volume    = {11},
  number    = {8},
  pages     = {e0160076},
  year      = {2016},
  publisher = {Public Library of Science},
  doi       = {10.1371/journal.pone.0160076}
}

@book{murray2007mathematical,
  title     = {Mathematical Biology: {I.} An Introduction},
  author    = {Murray, James D.},
  volume    = {17},
  edition   = {Third},
  year      = {2007},
  publisher = {Springer},
  isbn      = {978-0-387-95223-9}
}

@article{holling1959components,
  title     = {Some Characteristics of Simple Types of Predation and Parasitism},
  author    = {Holling, Crawford S.},
  journal   = {The Canadian Entomologist},
  volume    = {91},
  number    = {7},
  pages     = {385--398},
  year      = {1959},
  publisher = {Cambridge University Press},
  doi       = {}
}

@article{Cantrell1991,
  title     = {The Effects of Spatial Heterogeneity in Population Dynamics},
  author    = {Cantrell, Robert S. and Cosner, Chris},
  journal   = {Journal of Mathematical Biology},
  volume    = {29},
  pages     = {315--338},
  year      = {1991},
  doi       = {}
}

@article{Turnbull2007,
  title     = {How Spatial Structure Alters Population and Community Dynamics in a Natural Plant Community},
  author    = {Turnbull, Lindsay A. and Coomes, David A. and Purves, Drew W. and Rees, Mark},
  journal   = {Journal of Ecology},
  volume    = {95},
  number    = {1},
  pages     = {79--89},
  year      = {2007},
  publisher = {Wiley Online Library},
  doi       = {}
}

@article{Simoy2023,
  title     = {Non-Local Interaction Effects in Models of Interacting Populations},
  author    = {Simoy, Mario I. and Kuperman, Marcelo N.},
  journal   = {Chaos, Solitons and Fractals},
  volume    = {167},
  year      = {2023},
  pages     = {112993},
  publisher = {Elsevier},
  doi       = {10.1016/j.chaos.2022.112993}
}

@article{Rietkerk2021,
  title     = {Evasion of Tipping in Complex Systems through Spatial Pattern Formation},
  author    = {Rietkerk, Max and Bastiaansen, Robbin and Banerjee, Swarnendu and van de Koppel, Johan and Baudena, Mara and Doelman, Arjen},
  journal   = {Science},
  volume    = {374},
  number    = {6564},
  pages     = {eabj0359},
  year      = {2021},
  publisher = {American Association for the Advancement of Science},
  doi       = {10.1126/science.abj0359}
}

@article{Hastings1990,
  title     = {Spatial Heterogeneity and Ecological Models},
  author    = {Hastings, Alan},
  journal   = {Ecology},
  volume    = {71},
  number    = {2},
  pages     = {426--428},
  year      = {1990},
  publisher = {Wiley Online Library},
  doi       = {}
}

@article{Van_nes2005,
  title     = {Implications of Spatial Heterogeneity for Catastrophic Regime Shifts in Ecosystems},
  author    = {Van Nes, Egbert H. and Scheffer, Marten},
  journal   = {Ecology},
  volume    = {86},
  number    = {7},
  pages     = {1797--1807},
  year      = {2005},
  publisher = {Wiley Online Library},
  doi       = {10.1890/04-0550}
}

@article{Maciel2021,
  title     = {Enhanced Species Coexistence in {Lotka}-{Volterra} Competition Models Due to Nonlocal Interactions},
  author    = {Maciel, Gabriel A. and Martinez-Garcia, Ricardo},
  journal   = {Journal of Theoretical Biology},
  volume    = {530},
  pages     = {110872},
  year      = {2021},
  publisher = {Elsevier},
  doi       = {10.1016/j.jtbi.2021.110872}
}

@article{PintoRamos2025,
  title     = {Aperiodic Clustered and Periodic Hexagonal Vegetation Spot Arrays Explained by Inhomogeneous Environments and Climate Trends in Arid Ecosystems},
  author    = {Pinto-Ramos, David and Clerc, Marcel G. and Makhoute, Abdelkader and Tlidi, Mustapha},
  journal   = {Geophysical Research Letters},
  volume    = {52},
  number    = {21},
  pages     = {e2025GL118462},
  year      = {2025},
  publisher = {American Geophysical Union},
  doi       = {10.1029/2025GL118462}
}

@incollection{Schuessler2011,
  title     = {Evolution of the `Plant-Symbiotic' Fungal Phylum, Glomeromycota},
  author    = {Sch{\"u}{\ss}ler, Arthur and Walker, Christopher},
  booktitle = {Evolution of Fungi and Fungal-Like Organisms},
  series    = {The Mycota: A Comprehensive Treatise on Fungi as Experimental Systems for Basic and Applied Research},
  volume    = {14},
  editor    = {P{\"o}ggeler, Sandra and W{\"o}stemeyer, Joseph},
  publisher = {Springer},
  address   = {Dordrecht, Netherlands},
  pages     = {164--185},
  year      = {2011},
  isbn      = {978-3-642-19973-8},
  doi       = {10.1007/978-3-642-19974-5_7}
}

@article{Stanley2006,
  title     = {Photosymbiosis and the Evolution of Modern Coral Reefs},
  author    = {Stanley, George D.},
  journal   = {Science},
  volume    = {312},
  number    = {5775},
  pages     = {857--858},
  year      = {2006},
  publisher = {American Association for the Advancement of Science},
  doi       = {}
}

@article{Bascompte2007,
  title     = {Plant--Animal Mutualistic Networks: The Architecture of Biodiversity},
  author    = {Bascompte, Jordi and Jordano, Pedro},
  journal   = {Annual Review of Ecology, Evolution, and Systematics},
  volume    = {38},
  pages     = {567--593},
  year      = {2007},
  publisher = {Annual Reviews},
  doi       = {10.1146/annurev.ecolsys.38.091206.095818}
}

@article{Stone2020,
  title     = {The Stability of Mutualism},
  author    = {Stone, Lewi},
  journal   = {Nature Communications},
  volume    = {11},
  number    = {1},
  pages     = {2648},
  year      = {2020},
  publisher = {Nature Publishing Group},
  doi       = {10.1038/s41467-020-16474-4}
}

@article{AraujoLurgi2025,
  title     = {Mutualism Provides a Basis for Biodiversity in Eco-Evolutionary Community Assembly},
  author    = {Araujo, Gui and Lurgi, Miguel},
  journal   = {PLOS Computational Biology},
  volume    = {21},
  number    = {9},
  pages     = {e1013402},
  year      = {2025},
  publisher = {Public Library of Science},
  doi       = {10.1371/journal.pcbi.1013402}
}

@article{Bascompte2003,
  title     = {The Nested Assembly of Plant--Animal Mutualistic Networks},
  author    = {Bascompte, Jordi and Jordano, Pedro and Meli{\'a}n, Carlos J. and Olesen, Jens M.},
  journal   = {Proceedings of the National Academy of Sciences},
  volume    = {100},
  number    = {16},
  pages     = {9383--9387},
  year      = {2003},
  publisher = {National Academy of Sciences},
  doi       = {10.1073/pnas.1633576100}
}

@article{MartinezGarcia2020,
  title     = {How Range Residency and Long-Range Perception Change Encounter Rates},
  author    = {Martinez-Garcia, Ricardo and Fleming, Christen H. and Seppelt, Ralf and Fagan, William F. and Calabrese, Justin M.},
  journal   = {Journal of Theoretical Biology},
  volume    = {498},
  pages     = {110267},
  year      = {2020},
  publisher = {Elsevier},
  doi       = {10.1016/j.jtbi.2020.110267}
}

@article{Denk2024,
  title     = {Tipping Points Emerge from Weak Mutualism in Metacommunities},
  author    = {Denk, Jonas and Hallatschek, Oskar},
  journal   = {PLOS Computational Biology},
  volume    = {20},
  number    = {3},
  pages     = {e1011899},
  year      = {2024},
  publisher = {Public Library of Science},
  doi       = {10.1371/journal.pcbi.1011899}
}

@article{Sachs2006,
  title     = {Pathways to Mutualism Breakdown},
  author    = {Sachs, Joel L. and Simms, Ellen L.},
  journal   = {Trends in Ecology \& Evolution},
  volume    = {21},
  number    = {10},
  pages     = {585--592},
  year      = {2006},
  publisher = {Elsevier},
  doi       = {}
}

@article{toby2010mutualisms,
  title     = {Mutualisms in a Changing World: An Evolutionary Perspective},
  author    = {Kiers, E. Toby and Palmer, Todd M. and Ives, Anthony R. and Bruno, John F. and Bronstein, Judith L.},
  journal   = {Ecology Letters},
  volume    = {13},
  number    = {12},
  pages     = {1459--1474},
  year      = {2010},
  publisher = {Wiley Online Library},
  doi       = {}
}

@article{traveset2014mutualistic,
  title     = {Mutualistic Interactions and Biological Invasions},
  author    = {Traveset, Anna and Richardson, David M.},
  journal   = {Annual Review of Ecology, Evolution, and Systematics},
  volume    = {45},
  number    = {1},
  pages     = {89--113},
  year      = {2014},
  publisher = {Annual Reviews},
  doi       = {}
}

@article{Simard2012,
  title     = {Mycorrhizal Networks: Mechanisms, Ecology and Modelling},
  author    = {Simard, Suzanne W. and Beiler, Kevin J. and Bingham, Marcus A. and Deslippe, Julie R. and Philip, Leanne J. and Teste, Fran{\c{c}}ois P.},
  journal   = {Fungal Biology Reviews},
  volume    = {26},
  number    = {1},
  pages     = {39--60},
  year      = {2012},
  publisher = {Elsevier},
  doi       = {10.1016/j.fbr.2012.01.001}
}

@article{Detto2016,
  title     = {Stabilization of Species Coexistence in Spatial Models through the Aggregation--Segregation Effect Generated by Local Dispersal and Nonspecific Local Interactions},
  author    = {Detto, Matteo and Muller-Landau, Helene C.},
  journal   = {Theoretical Population Biology},
  volume    = {112},
  pages     = {97--108},
  year      = {2016},
  publisher = {Elsevier},
  doi       = {10.1016/j.tpb.2016.08.008}
}

@article{holland2010consumer,
  title     = {A Consumer--Resource Approach to the Density-Dependent Population Dynamics of Mutualism},
  author    = {Holland, J. Nathaniel and {DeAngelis}, Donald L.},
  journal   = {Ecology},
  volume    = {91},
  number    = {5},
  pages     = {1286--1295},
  year      = {2010},
  publisher = {Wiley Online Library},
  doi       = {}
}

@article{Michaels2024,
  title     = {When Patches Grow Themselves: From Analogy to Autocatalytic Processes, the Relevance of Ecological Nucleation for Restoration Practices},
  author    = {Michaels, Theo K. and Eppinga, Maarten B. and Bever, James D.},
  journal   = {Restoration Ecology},
  volume    = {32},
  number    = {1},
  pages     = {e14066},
  year      = {2024},
  publisher = {Wiley Online Library},
  doi       = {10.1111/rec.14066}
}

@book{samanta2021deterministic,
  title     = {Deterministic, Stochastic and Thermodynamic Modelling of Some Interacting Species},
  author    = {Samanta, Guruprasad},
  year      = {2021},
  publisher = {Springer},
  isbn      = {}
}

@article{deblauwe2008global,
  title     = {The Global Biogeography of Semi-Arid Periodic Vegetation Patterns},
  author    = {Deblauwe, Vincent and Barbier, Nicolas and Couteron, Pierre and Lejeune, Olivier and Bogaert, Jan},
  journal   = {Global Ecology and Biogeography},
  volume    = {17},
  number    = {6},
  pages     = {715--723},
  year      = {2008},
  publisher = {Wiley Online Library},
  doi       = {}
}

@article{stone2003pollination,
  title     = {Pollination Ecology of Acacias ({Fabaceae}, {Mimosoideae})},
  author    = {Stone, Graham N. and Raine, Nigel E. and Prescott, Matthew and Willmer, Pat G.},
  journal   = {Australian Systematic Botany},
  volume    = {16},
  number    = {1},
  pages     = {103--118},
  year      = {2003},
  publisher = {CSIRO Publishing},
  doi       = {}
}

@article{golubov1999demography,
  title     = {Demography of the Invasive Woody Perennial {Prosopis} juliflora (Giant Mesquite) in a Semi-Desert of {Mexico}},
  author    = {Golubov, Jordan and Mandujano, Maria C. and Monta{\~n}a, Carlos and L{\'o}pez-Portillo, Jorge and Eguiarte, Luis E.},
  journal   = {Journal of Applied Ecology},
  volume    = {36},
  number    = {6},
  pages     = {925--935},
  year      = {1999},
  publisher = {Wiley Online Library},
  doi       = {}
}

@book{boffa1999agroforestry,
  title     = {Agroforestry Parklands in Sub-{Saharan} {Africa}},
  author    = {Boffa, Jean-Marc},
  volume    = {34},
  year      = {1999},
  publisher = {Food \& Agriculture Organization},
  isbn      = {}
}

@article{Akhtar2026,
  title     = {Study on Fear-Induced Group Defense and Cooperative Hunting: Bifurcation, Uncertainty, Seasonality, and Spatio-Temporal Analysis in Predator--Prey System},
  author    = {Akhtar, Parvez and Santra, Nirapada and Samanta, Guruprasad},
  journal   = {The European Physical Journal Special Topics},
  pages     = {1--40},
  year      = {2026},
  publisher = {Springer},
  doi       = {10.1140/epjs/s11734-026-02123-2}
}

@article{PintoRamos2026,
  title     = {How Spatial Patterns Can Lead to Less Resilient Ecosystems},
  author    = {Pinto-Ramos, David and Martinez-Garcia, Ricardo},
  journal   = {Proceedings of the National Academy of Sciences},
  volume    = {123},
  number    = {14},
  pages     = {e2511994123},
  year      = {2026},
  publisher = {National Academy of Sciences},
  doi       = {}
}

@article{Godoy2026,
  title     = {Cooperation Maximizes Biodiversity},
  author    = {Godoy, Oscar and Soler-Toscano, Fernando and Portillo, Jos{\'e} R. and Su{\'a}rez, Antonio and Langa, Jos{\'e} A.},
  journal   = {bioRxiv},
  year      = {2024},
  doi       = {10.1101/2024.10.22.619656}
}

@article{Menezes2025,
  title     = {The Range-Resident Logistic Model: A New Framework to Formalise the Population-Dynamics Consequences of Range Residency},
  author    = {Menezes, Rafael and Calabrese, Justin M. and Fagan, William F. and Prado, Paulo In{\'a}cio and Martinez-Garcia, Ricardo},
  journal   = {Ecology Letters},
  volume    = {28},
  number    = {12},
  pages     = {e70269},
  year      = {2025},
  publisher = {Wiley Online Library},
  doi       = {}
}

@article{Plank2025,
  title     = {Random Walk Models in the Life Sciences: Including Births, Deaths and Local Interactions},
  author    = {Plank, Michael J. and Simpson, Matthew J. and Baker, Ruth E.},
  journal   = {Journal of the Royal Society Interface},
  volume    = {22},
  number    = {222},
  pages     = {20240422},
  year      = {2025},
  publisher = {The Royal Society},
  doi       = {}
}

@article{MartinezGarcia2023,
  title     = {Integrating Theory and Experiments to Link Local Mechanisms and Ecosystem-Level Consequences of Vegetation Patterns in Drylands},
  author    = {Martinez-Garcia, Ricardo and Cabal, Ciro and Calabrese, Justin M. and Hern{\'a}ndez-Garc{\'i}a, Emilio and Tarnita, Corina E. and L{\'o}pez, Crist{\'o}bal and Bonachela, Juan A.},
  journal   = {Chaos, Solitons \& Fractals},
  volume    = {166},
  pages     = {112881},
  year      = {2023},
  publisher = {Elsevier},
  doi       = {}
}

@article{Jelle2026,
  title     = {Vegetation Pattern Formation Induced by Local Growth Outpacing Susceptibility to Non-Local Competition},
  author    = {Van Der Voort, Jelle and Martinez-Garcia, Ricardo and Doelman, Arjen},
  journal   = {Journal of Theoretical Biology},
  volume    = {626},
  pages     = {112441},
  year      = {2026},
  publisher = {Elsevier},
  doi       = {10.1016/j.jtbi.2026.112441}
}

@article{Kastner2025,
  title     = {Formation of Spatial Vegetation Patterns in Heterogeneous Environments},
  author    = {K{\"a}stner, Karl and Caviedes-Voull{\`e}me, Daniel and Hinz, Christoph},
  journal   = {PLOS One},
  volume    = {20},
  number    = {5},
  pages     = {e0324181},
  year      = {2025},
  publisher = {Public Library of Science},
  doi       = {10.1371/journal.pone.0324181}
}

@article{Kastner2024,
  title     = {Unravelling the Spatial Structure of Regular Dryland Vegetation Patterns},
  author    = {K{\"a}stner, Karl and {Van De Vijsel}, Roeland C. and Caviedes-Voull{\`e}me, Daniel and Frechen, Nanu T. and Hinz, Christoph},
  journal   = {CATENA},
  volume    = {247},
  pages     = {108442},
  year      = {2024},
  publisher = {Elsevier},
  doi       = {10.1016/j.catena.2024.108442}
}

@article{PintoRamos2026global,
  title     = {Global Remote Sensing Reveals Vegetation Clustering as a Physical Footprint of Shifting Aridity Trends in Drylands},
  author    = {Pinto-Ramos, David and Clerc, Marcel G. and Makhoute, Abdelkader and Tlidi, Mustapha},
  journal   = {arXiv preprint arXiv:2604.22122},
  year      = {2026},
  doi       = {}
}

@article{Lefever1997,
  title     = {On the Origin of Tiger Bush},
  author    = {Lefever, Ren{\'e} and Lejeune, Olivier},
  journal   = {Bulletin of Mathematical Biology},
  volume    = {59},
  number    = {2},
  pages     = {263--294},
  year      = {1997},
  publisher = {Springer},
  doi       = {}
}

@article{Hernandez-Garcia2004,
  title     = {Clustering, Advection, and Patterns in a Model of Population Dynamics with Neighborhood-Dependent Rates},
  author    = {Hern{\'a}ndez-Garc{\'i}a, Emilio and L{\'o}pez, Crist{\'o}bal},
  journal   = {Physical Review E},
  volume    = {70},
  number    = {1},
  pages     = {016216},
  year      = {2004},
  publisher = {American Physical Society},
  doi       = {}
}

@article{Piva2021,
  title     = {Interplay between Scales in the Nonlocal {FKPP} Equation},
  author    = {Piva, G. G. and Colombo, Eduardo H. and Anteneodo, C.},
  journal   = {Chaos, Solitons and Fractals},
  volume    = {153},
  pages     = {111609},
  year      = {2021},
  publisher = {Elsevier},
  doi       = {10.1016/j.chaos.2021.111609}
}

@article{Cabal2020,
  title     = {The Exploitative Segregation of Plant Roots},
  author    = {Cabal, Ciro and Martinez-Garcia, Ricardo and {De Castro}, A. and Valladares, F. and Pacala, Stephen W.},
  journal   = {Science},
  volume    = {370},
  number    = {6251},
  pages     = {1197--1199},
  year      = {2020},
  publisher = {American Association for the Advancement of Science},
  doi       = {10.1126/science.aba9877}
}

@article{Cabal2024,
  title     = {Plant Antagonistic Facilitation across Environmental Gradients: A Soil-Resource Ecosystem Engineering Model},
  author    = {Cabal, Ciro and Maciel, Gabriel A. and Martinez-Garcia, Ricardo},
  journal   = {New Phytologist},
  volume    = {244},
  number    = {2},
  pages     = {670--682},
  year      = {2024},
  publisher = {Wiley Online Library},
  doi       = {}
}

@article{Surendran2025,
  title     = {Spatial Moment Dynamics and Biomass Density Equations Provide Complementary, yet Limited, Descriptions of Pattern Formation in Individual-Based Simulations},
  author    = {Surendran, Anudeep and Pinto-Ramos, David and Menezes, Rafael and Martinez-Garcia, Ricardo},
  journal   = {Physica D: Nonlinear Phenomena},
  volume    = {477},
  pages     = {134703},
  year      = {2025},
  publisher = {Elsevier},
  doi       = {}
}

@article{Lopez2004,
  title     = {Fluctuations Impact on a Pattern-Forming Model of Population Dynamics with Non-Local Interactions},
  author    = {L{\'o}pez, Crist{\'o}bal and Hern{\'a}ndez-Garc{\'i}a, Emilio},
  journal   = {Physica D: Nonlinear Phenomena},
  volume    = {199},
  number    = {1-2},
  pages     = {223--234},
  year      = {2004},
  publisher = {Elsevier},
  doi       = {}
}
\vspace{1cm}
\setcounter{figure}{0}      
\renewcommand\thefigure{A\arabic{figure}}

\section*{Appendices}
\begin{appendices}
\section{Linear stability analysis of the spatial model}
\label{ap:A}

We start from the dimensionless version of the model equations \eqref{eq:spatialveg}-\eqref{eq:spatialpol} where we leave the establishment probability implicit
    \begin{align}
        \dfrac{\partial v(x,t)}{\partial t} =& ( 1 - v)vpP_e(\tilde{v}, \alpha) - v + \overline{D}_v\frac{\partial^2 v}{\partial x^2}, \\
        \dfrac{\partial p(x,t)}{\partial t} =& \beta p - \gamma p^2 + \mu vp + \overline{D}_p\frac{\partial^2 p}{\partial x^2},
    \end{align}
and introduce small perturbations in each of the fields around the fixed points, $p(x,t)=p^*+\delta p$ and $v(x,t)=v^*+\delta v$. Then, we obtain the linear equations for the dynamics of the perturbations
    \begin{align}
        \dfrac{\partial (\delta v(x,t))}{\partial t} =& p^*(1-2v^*)P_e(\tilde{v}^*, \alpha)\delta v + v^*(1-v^*)P_e(\tilde{v}^*, \alpha)\delta p\; +\\
        & (1-v^*)v^*p^* \dfrac{\partial}{\partial \tilde{v}}P_e(\tilde{v}^*, \alpha)(\delta v \ast G) - \delta v + \overline{D}_v\frac{\partial^2 \delta v}{\partial x^2}, \\
        \dfrac{\partial (\delta p(x,t))}{\partial t} =& \mu p^*\delta v + (\beta - 2\gamma p^* + \mu v^*)\delta p + \overline{D}_p\frac{\partial^2 \delta p}{\partial x^2},
    \end{align}
where we have defined
\begin{eqnarray}
    (\delta v \ast G)= \int \delta v(x')G(x-x')dx',
\end{eqnarray}
and $G(x-x')$ is the top-hat kernel, that is
\begin{eqnarray}
    G(x-x')= \begin{cases} \frac{1}{2} \;\; \text{if} \;\; |x-x'|<1, \\
    0\;\; \text{otherwise}.
    \end{cases}
\end{eqnarray}
Then, we determine the linear growth-rate, $\lambda({k})$, of the Fourier modes $\widehat{\delta v}({k}, t)$, defined from the Fourier transform
$$\delta v(x,t) = \int \widehat{\delta v}({k}, t) e^{ i {k} x} d {k}.$$
Inserting this in our linearized equations leads to
    \begin{align}
        \dfrac{\partial (\widehat{\delta v}({k},t))}{\partial t} =& \left[p^*(1-2v^*)P_e(\tilde{v}^*, \alpha) + (1-v^*)v^*p^* \dfrac{\partial}{\partial \tilde{v}}P_e(\tilde{v}^*, \alpha)\hat{G}({k}) - 1\right]\widehat{\delta v}\;+ \nonumber \\
        & v^*(1-v^*)P_e(\tilde{v}^*, \alpha)\widehat{\delta p} - \overline{D}_v {k}^2 \widehat{\delta v}, \nonumber \\
        \dfrac{\partial (\widehat{\delta p}({k},t))}{\partial t} =& \mu p^*\widehat{\delta v} + (\beta - 2\gamma p^* + \mu v^*)\widehat{\delta p} - \overline{D}_p {k}^2\widehat{\delta p},
        \label{eq:fourier}
    \end{align}
where $\hat{G}({k})=\sin({k})/{k}$ is the Fourier transform of the step function kernel. The Eq.\,\eqref{eq:fourier} is a linear equation in the Fourier mode with solutions of the form
$$ \begin{pmatrix}
    \widehat{\delta v}\\
    \widehat{\delta p}
\end{pmatrix}\propto e^{\lambda({k})t}.$$
Spatial patterns can form when at least one nonzero wavenumber ${k}$ satisfies $\lambda({k})>0$. A positive growth rate causes the associated Fourier mode to grow exponentially in the linear regime and potentially lead to a spatial pattern. The functions $\lambda({k})$ are the eigenvalues of the eigenvalue problem defined by Eq.\,\eqref{eq:fourier}. This means that the determinant of $A-\lambda I$ vanishes, where
\begin{eqnarray}
        A = \begin{pmatrix}
        p(1-2v)\dfrac{1}{1+\alpha v} - \dfrac{\alpha (1-v)vp}{(1+\alpha v)^2}\hat{G}({k}) - 1 - \overline{D}_v {k}^2 \; & \; v(1-v)\dfrac{1}{1+\alpha v} \\
        \mu p & \beta - 2\gamma p + \mu v - \overline{D}_p {k}^2
    \end{pmatrix}_{(v,p)^*}.
\end{eqnarray}
Then. the growth rates are defined in terms of the elements of $A$ as    \begin{equation}
        \lambda_{1,2}({k}) = \dfrac{1}{2}\left(\mathrm{tr}(A) \pm \sqrt{\mathrm{tr}^2(A) - 4\mathrm{det}(A)}\right),
    \end{equation}
where $\text{tr}(\cdot)$ and $\text{det}(\cdot)$ denote the trace and the determinant, respectively.

\end{appendices}

\end{document}